\documentclass[preprint,superscriptaddress,nofootinbib]{revtex4-1}

\usepackage{amsmath,amssymb,mathtools}
\usepackage{bm}
\usepackage{slashed}
\usepackage{ulem}
\usepackage{url}
\usepackage{color}
\usepackage[bookmarks=true,bookmarksnumbered=true,bookmarkstype=toc, colorlinks=true, citecolor=blue, linkcolor=blue ,urlcolor=blue]{hyperref}

\renewcommand{\thesubsection}{\arabic{section}.\arabic{subsection}}

\numberwithin{equation}{section}

\definecolor{fgreen}{cmyk}{0.91,0,0.88,0.12}

\definecolor{pink}{cmyk}{0,0.5,0,0}

\definecolor{pastelpink}{cmyk}{0,0.25,0,0}

\definecolor{softpink}{cmyk}{0,0.125,0,0}

\definecolor{purple}{cmyk}{0.5,1.0,0.1,0}

\definecolor{violet}{cmyk}{0.75,1,0.25,0}

\newcommand\dd{\mathrm{d}}

\begin{document}

\preprint{UME-PP-026}
\preprint{KYUSHU-HET-262}

\title{
New Constraint on Dark Photon at T2K Off-Axis Near Detector
}

\author{Takeshi Araki}
\email{t-araki@den.ohu-u.ac.jp}
\affiliation{Faculty of Dentistry, Ohu University, 31-1 Misumidou, Tomita-machi, Koriyama, Fukushima 963-8611,  Japan}

\author{Kento Asai}
\email{kento@icrr.u-tokyo.ac.jp}
\affiliation{Institute for Cosmic Ray Research (ICRR), The University of Tokyo, Kashiwa, Chiba 277--8582, Japan}

\author{Tomoya Iizawa}
\email{Tomoya.Iizawa@cern.ch}
\affiliation{Institute of Particle and Nuclear Studies, High Energy Accelerator Research Organization (KEK), Oho 1-1, Tsukuba, Ibaraki 305-0801, Japan}

\author{\\Hidetoshi Otono}
\email{otono@phys.kyushu-u.ac.jp}
\affiliation{Research Center for Advanced Particle Physics, Kyushu University, 744 Motooka, Nishi-ku, Fukuoka, 819–0395, Japan}

\author{Takashi Shimomura}
\email{shimomura@cc.miyazaki-u.ac.jp}
\affiliation{Faculty of Education, University of Miyazaki, 1-1 Gakuen-Kibanadai-Nishi, Miyazaki 889-2192, Japan}
\affiliation{Department of Physics, Kyushu University, 744 Motooka, Nishi-ku, Fukuoka, 819–0395, Japan}

\author{Yosuke Takubo}
\email{yosuke.takubo@kek.jp}
\affiliation{Institute of Particle and Nuclear Studies, High Energy Accelerator Research Organization (KEK), Oho 1-1, Tsukuba, Ibaraki 305-0801, Japan}

\begin{abstract}
The T2K experiment is one of the most powerful long-baseline experiments to investigate neutrino oscillations. The off-axis near detector called ND280 is installed 
280\,m downstream from the neutrino production target to measure the neutrino energy spectrum.
In this paper, we study the capability of the ND280 detector to search for the dark photon produced through the meson rare decay and proton bremsstrahlung processes at the proton beam dump. 
We find that the ten-year operation of T2K with the ND280 detector excludes the unexplored parameter region for the dark photon mass and kinetic mixing.
We also show that a broader parameter region can be searched by the ND280 in the future T2K operation for dark photon as well as U(1)$_{B-L}$ gauge boson.
\end{abstract} 


\maketitle

\section{Introduction}
\label{sec:introduction}
No direct observation of the dark matter except through gravity leads us to the idea of dark sectors. 
The dark sector consists of new particles including the dark matter and is sequestered from the Standard Model (SM) sector. 
It also contains so-called portal particles which weakly connect two sectors. 
Among such portals, the dark photon is a massive hypothetical vector boson that mixes with the electromagnetic photon. 
Extensive searches have already placed stringent bounds on its interaction strength and mass. 
Further searches continue at ongoing and planned experiments with higher statistics and low-background detectors.

The dark photon is in general defined by a gauge boson of a new Abelian symmetry under which the SM particles are uncharged. 
In such models, there are no direct interactions of the dark photon with the SM matters, except for the kinetic mixing with the electromagnetic photon (or hypercharge gauge boson) \cite{Okun:1982xi,Galison:1983pa,Holdom:1985ag,Foot:1991kb,Babu:1997st}. 
Due to the mixing, the dark photon can interact with the SM matters through the electromagnetic current. 
In this sense, it is called ``dark photon" \cite{Pospelov:2007mp,Huh:2007zw,Pospelov:2008zw}. 
A light dark photon with mass below $1$\,GeV has been particularly explored in the decays of light mesons $(\pi,~\eta,~\eta')$ and bremsstrahlung in electron and proton beam dump experiments. 
The dark photon can be produced by replacing a photon appearing in these processes. 
Null results from these experiments excluded the kinetic mixing parameter roughly between $10^{-8}$ and $10^{-4}$ (see refs.~\cite{Bauer:2018onh,Fabbrichesi:2020wbt,Caputo:2021eaa} for recent reviews and references). 
Below the lower bound, the dark photon becomes so feebly interacting and long-lived particle. 
In order to search for such particles, one needs higher statistics and low-background detectors placed far away from the production point. 
Several experiments like NA62 \cite{NA62:2017rwk, Lanfranchi:2017wzl} and FASER \cite{Feng:2017uoz, FASER:2022hcn, FASER:2021cpr, FASER:2021ljd} are running, and there are also proposals for new experiments: for instance   
CODEX-b \cite{Gligorov:2017nwh,Aielli:2019ivi}, 
FACET \cite{Cerci:2021nlb}, 
MATHUSLA \cite{Chou:2016lxi,Curtin:2018mvb}, 
SeaQuest \cite{Berlin:2018pwi,Batell:2020vqn}, and
SHiP \cite{Alekhin:2015byh,SHiP:2015vad}.
Notably, the NA62 and the FASER collaboration have recently released their first results on the dark photon search \cite{na62collaboration2023search, NA62LaThuile, CERN-FASER-CONF-2023-001}, which excluded new parameter space.

The Tokai-to-Kamioka (T2K) experiment \cite{T2K:2011qtm} can offer a new opportunity to search for the dark photon. 
The T2K experiment is a long-baseline neutrino oscillation experiment operated since 2010 and will be upgraded by increasing its beam power in the next several years. 
The proton beams accelerated up to $30$\,GeV at J-PARC strike graphite targets, and produce a large number of neutral mesons as well as charged ones. 
The dark photons can be produced through the decays of these neutral mesons and bremsstrahlung of the proton beams.
The ND280 off-axis detector is expected to be suitable in searches for the dark photon or long-lived particles decaying into charged particles.\footnote{
There is also the INGRID on-axis detector installed at 280\,m from the target. The INGRID has less tracking performance of charged particles without a magnet system and with a shorter tracking volume with respect to the ND280. Thus, we do not consider this in our study.
}  
It is located 280\,m away from the neutrino production target and shielded by the ground.
Thus, backgrounds originated from mesons, photons, and muons will be reduced significantly. 
In refs. \cite{Asaka:2012bb,T2K:2019jwa} and \cite{Gorbunov:2021jog}, the searches for heavy neutral leptons and millicharged particles were studied for ND280. 
In this work, we investigate the capability of dark photon search with the ND280 detector at the T2K experiment, deriving a new exclusion limit on its mass and kinetic mixing. 
We will show that the ten-year operation of T2K with the ND280 detector excludes the unexplored parameter region for the dark photon mass and kinetic mixing.
We also present that a broader parameter region can be searched by the ND280 in the future T2K operation for dark photon as well as U(1)$_{B-L}$ gauge boson.

This paper is organized as follows. 
In section~\ref{sec:model}, we describe the minimal dark photon model and show their production and decays. 
In section~\ref{sec:beam}, the simulation for the production of the secondary particles in the target at the T2K proton beam line is described, then the ND280 off-axis detector is explained in section~\ref{sec:detector}. 
In section~\ref{sec:result}, we present new constraints and future sensitivity on the mass and kinetic mixing of the dark photon and U(1)$_{B-L}$ gauge boson derived from our analyses. 
Section~\ref{sec:summary} is devoted to summary.

\section{Dark Photon Model}
\label{sec:model}

\subsection{Lagrangian}
We extend the SM by adding a new Abelian gauge symmetry U(1)$'$ under which all the SM particles are neutral.
In this case, the SM Lagrangian is augmented by only two kinetic terms
\begin{align}
{\cal L}_{\rm kinetic} =  
- \frac{1}{4} F'_{\mu\nu} F'^{\mu\nu}
- \frac{\varepsilon'}{2} F'_{\mu\nu} F^{\mu\nu} ~,
\end{align}
where $F_{\mu\nu}$ and $F'_{\mu\nu}$ are the field strength tensors of the SM U(1) hypercharge symmetry and U(1)$'$, respectively.
The second term is the so-called kinetic mixing term, and it causes interaction between the U(1)$'$ gauge boson and the SM particles.

In the diagonal basis of the gauge kinetic terms, we consider the following Lagrangian:
\begin{align}
{\cal L}_{\rm DP}^{} = 
- \varepsilon e A_{\mu} ' J_{\rm EM}^{\mu}
+ \frac{1}{2}m_{A'}^{2} A_\mu' A'^{\mu} ~,
\label{eq:Ldp}
\end{align}
where $e$ is the electric charge, and $J_{\rm EM}^\mu$ is the electromagnetic current of the SM.
The new gauge boson associated with U(1)$'$ is denoted as $A'_\mu$, that is to say, $A'_\mu$ stands for the dark photon.
Note that we assume U(1)$'$ is broken, and thus $A'_\mu$ has a non-zero mass $m_{A'}^{}$. 
In the first term, we define $\varepsilon \equiv \varepsilon' \cos\theta_{\rm w}$, where $\theta_{\rm w}$ is the weak mixing angle, and will refer to $\varepsilon$ as the kinetic mixing parameter in what follows.

\subsection{Production of dark photon}
In experiments with high energy proton beam, such as T2K, dark photons are copiously produced from a variety of production channels.
For instance, a meson decaying into the SM photons, $\gamma$, can be an important source of dark photons, because $\gamma$ can be replaced by $A'$ through the kinetic mixing.
Among them, dominant production modes are light neutral meson decays, such as $\pi^0 \rightarrow A'\gamma$, $\eta \rightarrow A'\gamma$, $\eta' \rightarrow A'\gamma$, and $\eta' \rightarrow A'\rho$, which produce $A'$ up to $m_{A'} < 0.5$\,GeV.
The branching ratios of these decays are given by \cite{Batell:2009di}
\begin{align}
& {\rm BR}(X \rightarrow A'\gamma)
= 2 \varepsilon^2 \left( 1-\frac{m_{A'}^2}{m_{X}^2} \right)^3 {\rm BR}(X \rightarrow \gamma\gamma)~, 
\label{eq:BRprod}
\end{align}
where $X=\pi^0,~\eta$, or $\eta'$, and 
\begin{align}
& {\rm BR}(\eta' \rightarrow A'\rho)
= \varepsilon^2 
~\frac{
\big[
(m_{A'}^2 - (m_{\eta'}+m_{\rho})^2)
(m_{A'}^2 - (m_{\eta'}-m_{\rho})^2)
\big]^{3/2}
}
{(m_{\eta'}^2-m_{\rho}^2)^{3}}
~{\rm BR}(\eta' \rightarrow \gamma\rho)~.
\end{align}
In our numerical calculations, we use 
${\rm BR}(\pi^0 \rightarrow \gamma\gamma)=0.99$, 
${\rm BR}(\eta \rightarrow \gamma\gamma)=0.39$, 
${\rm BR}(\eta' \rightarrow \gamma\gamma)=0.023$, and 
${\rm BR}(\eta' \rightarrow \gamma\rho)=0.30$ \cite{ParticleDataGroup:2022pth}.
 
In addition to the meson decays, dark photons can be also produced from proton bremsstrahlung.
In particular, for 0.5\,GeV $< m_{A'}^{} <$ 1\,GeV, the production rate could be resonantly enhanced via the mixing with $\rho$ and $\omega$ mesons, and 
hence a large amount of $A'$ is expected in that mass range.
The exact calculation of the bremsstrahlung process, however, is difficult due to nonperturbative QCD effects.
In order to estimate the number of events, we rely on the Weizs\"{a}cker-Williams (WW) approximation~\cite{Fermi:1924tc,vonWeizsacker:1934nji,Williams:1934ad,Kim:1973he,Blumlein:2011mv,Blumlein:2013cua} and follow the calculation procedure outlined in ref.~\cite{Feng:2017uoz}.
We will discuss in more detail in section~\ref{sec:result}.

\subsection{Decay of dark photon}
Through the kinetic mixing, dark photons can decay into the SM particles if kinematically allowed.
For decays into leptons, the partial decay widths are given by
\begin{align}
\Gamma(A' \rightarrow \ell\overline{\ell}) = 
\frac{\varepsilon^2 e^2}{12\pi} m_{A'}^{}
\left( 1+\frac{2 m_\ell^2}{m_{A'}^2} \right)
\sqrt{ 1-\frac{4 m_\ell^2}{m_{A'}^2} } ~,
\label{eq:Gll}
\end{align}
where $\ell = e,\mu,\tau$.
The decay widths for hadronic final states can be calculated by using the hadronic production cross section measured at $e^+ e^-$ colliders, such as
\begin{align}
\Gamma(A' \rightarrow {\rm hadrons}) =
\Gamma(A' \rightarrow \mu\overline{\mu}) ~ R(m_{A'}^2) ~,
\end{align}
where $R(s) = \sigma(e^+ e^- \rightarrow {\rm hadrons}) / \sigma(e^+ e^- \rightarrow \mu^+ \mu^-)$, and $s$ is the center of mass energy.
In our numerical calculations, nevertheless, we make use of the date files included in the FORESEE package~\cite{Kling:2021fwx}.

\section{T2K Beam Simulation}
\label{sec:beam}

GEANT4 package~\cite{GEANT4:2002zbu} version 4.11.0.3 with physics package of QGSP\_BERT is used to generate the primary protons and simulate interactions of particles in the materials. 
The protons with 30\,GeV are injected to the graphite target with 13~mm in radius and 90~cm in length. 
The origin of the coordinate is defined to be at the center of the graphite target in the $x$-$y$ plane and at the upstream end of the graphite target in the $z$-coordinate which is parallel to the proton beam direction. 

Three Electro-Magnetic Horns (EMHorns) with two layers of a simplified shape are placed right after the target to focus secondary charged particles. The radii of the inner (outer) layers are set to 27 (200), 40 (500), and 70 (700)~mm for the first, second, and third EMHorns, respectively, with 3~mm thickness of the aluminum. Their lengths and central positions are set to (1.5\,m, 0.75\,m), (2.0\,m, 2.81\,m), and (2.5\,m, 9.98\,m), respectively. The magnetic field is applied with a formula of $B[\mathrm{T}] = 0.063/r\,[\mathrm{m}]$ to reproduce that shown in Fig.~7 in ref.~\cite{T2K:2012bge}, where $r$ is a radius from the $z$-axis.


Injecting $10^{5}$ protons on the target, the momentum and angle with respect to the beam axis distributions are obtained for $\pi^{0}$, $\eta$, and $\eta^{\prime}$, that are generated from interactions of the protons in the target. 
They are used as the initial condition to simulate production of the dark photons from the decays of these particles.


\section{T2K Near Detectors}
\label{sec:detector}

The ND280 detector is designed to measure the neutrino energy spectrum with neutral and charged current interactions, and is placed at 280\,m from the neutrino production target and $2.5^{\circ}$ away from the beam axis. 
It consists of the Pi-zero Detector (P{\O}D) \cite{Assylbekov:2011sh}, Fine Grained Detectors (FGDs)~\cite{T2KND280FGD:2012umz}, Time Projection Chambers (TPCs)~\cite{T2KND280TPC:2010nnd}, Electromagnetic CALorimeter (ECAL)~\cite{T2KUK:2013wkh} and Side Muon Range Detector (SMRD)~\cite{Aoki:2013swe}, that are installed in the previous UA1 magnet operated at 0.2\,T. 
The P{\O}D is a scintillator based tracking calorimeter optimized to measure $\pi^{0}$ produced in the neutral current interactions of neutrinos. 
The FGDs consist of two layers of scintillator bars and are placed between three TPC layers. 
The charged particles created in the charged current interactions are measured with FGDs and TPCs. 
The ECAL is placed to surround P{\O}D, FGDs and TPCs, and the SMRD is integrated inside the return yoke of the magnet.

In T2K-II starting from 2023, P{\O}D is replaced by three new detectors, Super-Fine Grained Detector (SuperFGD)~\cite{Dergacheva:2022omv}, High Angle (HA) TPC~\cite{Attie:2023ttw} and Time Of Flight (ToF)~\cite{Korzenev:2019kud}. 
SuperFGD consists of 2.1 million scintillator cubes with $1 \times 1 \times 1$\,cm$^{3}$, that are traversed by three Wave Length Shifting (WLS) fibers. 
The total size of SuperFGD is $192 \times 182 \times 56$\,cm$^{3}$. 
HA-TPC is placed bellow and above the SuperFGD. 
It consists of two TPC layers with resistive Micromegas modules, and its total size is $181 \times 223 \times 85$\,cm$^{3}$. 
The ToF system surrounds the SuperFGD and high angle TPCs, and consists of six panels with $250 \times 230$\,cm$^{2}$ which is comprised of 20 scintillator bars. 

In this study, we assume the region of TPCs in the ND280 at T2K-II as the effective area, which is 82\% of the total volume. For the signal selection, we require $\Delta\Phi < 90^{\circ}$ and $\cos\theta > 0.992$, where $\Delta\Phi$ is angle between two charged tracks in the final states and $\theta$ is polar angle of the reconstructed dark photon. These requirements follow those used in search for heavy neutrinos in the final states of $\mu^{\pm}\pi^{\mp}$ or $e^{\pm}\pi^{\mp}$ with the ND280 at T2K \cite{T2K:2019jwa}, except for ignoring the selection cut of the invariant mass less than 700 MeV. 
We assume 0.25 of the signal selection efficiency by referring to Fig. 4 
in ref.~\cite{T2K:2019jwa} while it will be improved with ND280 at T2K-II due to the 
better track reconstruction efficiency.


\section{Numerical calculations and results}
\label{sec:result}

\subsection{Number of signal events}
As we mentioned in section \ref{sec:model}, we consider two types of production modes of dark photons, i.e., via the meson decays and proton bremsstrahlung.
Thus, the number of signal events is divided into two parts and given by
\begin{align}
N_{\rm sig} 
= \int (~\dd N_{\rm DP}^{\rm meson} + \dd N_{\rm DP}^{\rm brems}~)
\times 
{\cal P}^{\rm det}(|{\bm p}_{A'}^{}|, \theta_{A'})
\times {\rm BR}(A' \rightarrow f\bar{f})
\times {\rm EFCY}
\times 0.82~.
\label{eq:Nsig}
\end{align}
In the parentheses, the first term, $\dd N_{\rm DP}^{\rm meson}$, stands for the number of dark photons produced from the meson decays, and it is written by
\begin{align}
\dd N_{\rm DP}^{\rm meson} =
N_{\rm pot} \sum_{X, Y}~
\dd|{\bm p}_X^{}|~ \dd\theta_X~
\frac{\dd^2 N_X}{\dd|{\bm p}_X^{}| \dd\theta_X}~
{\rm BR}(X \rightarrow A' Y)~,
\label{eq:Nmeson}
\end{align}
where $N_{\rm pot}$ is protons on target (POT).
The momentum of a meson $X$, its angle with respect to the beam axis, and the differential flux are denoted by ${\bm p}_{X}^{}$, $\theta_{X}$ and $\dd^2 N_X/\dd|{\bm p}_{X}^{}| \dd\theta_X$, respectively, and they are numerically generated as explained in section~\ref{sec:beam}.
We incorporate productions from $X = \pi^0,\eta$ mesons with $Y=\gamma$, and 
$X=\eta^\prime$ with $Y=\gamma$ and $\rho$.
The overall factor ${\cal P}^{\rm det}(|{\bm p}_{A'}^{}|, \theta_{A'})$ in eq. (\ref{eq:Nsig}) is a probability that a dark photon having a momentum ${\bm p}_{A'}^{}$ and an angle $\theta_{A'}$ decays inside the detector, and it is given in eq.~\eqref{eq:Paprx}.
Note that ${\bm p}_{A'}^{}$ is determined by $|{\bm p}_{X}^{}|$, $\theta_{X}$ and the angles of the dark photon momentum; Those angles are generated and integrated out by running Monte Carlo simulations.
For the dark photon branching ratio ${\rm BR}(A' \rightarrow f\bar{f})$, in this work, we take into account $f=e^{\pm},\mu^{\pm},\pi^{\pm}$ with the charged track reconstruction efficiency of EFCY~=~0.25.
Lastly, in order to take into account the signal events only in TPCs, the factor 0.82 (82\%) is multiplied.

As for the bremsstrahlung production, we follow the calculation procedure summarized in ref.~\cite{Feng:2017uoz}\footnote{
We follow the appendix B in the preprint version 3 of ref.~\cite{Feng:2017uoz}.
} and define the number of produced dark photons as 
\begin{align}
\dd N_{\rm DP}^{\rm brems} =
N_{\rm pot} |F(m_{A'}^2)|^2~
\dd z~ \dd p_{A',t}^2 ~
\frac{\sigma_{pp}(s')}{\sigma_{pp}(s)}~
w(z, p_{A',t}^2) ~ \Theta(\Lambda^2_{\rm QCD}-q^2_{\rm min})~,
\label{eq:Nbrmss}
\end{align}
where $z=p_{A',z}^{}/P_{\rm beam}$ with $P_{\rm beam}$ being the beam momentum, $p_{A',t}^{}$ is the transverse momentum of a dark photon, $w(z, p_{A',t}^2)$ is the splitting function derived in ref.~\cite{Blumlein:2013cua}\footnote{
In ref.~\cite{Gorbunov:2023jnx}, the validity of the splitting function is studied by taking into account the non-zero momentum transfer between protons for the case of {\it elastic} proton scattering.
}
, and $\Theta(\Lambda_{\rm QCD}^2 - q_{\rm min}^2)$ is a Heaviside step function to ensure the validity of the WW approximation.
As for the proton-proton inelastic cross section, $\sigma_{pp}$, we read off the data provided by the particle data group~\cite{ParticleDataGroup:2022pth}.
Note that the probability ${\cal P}^{\rm det}(|{\bm p}_{A'}^{}|, \theta_{A'})$ is integrated over $z$ and $p_{A',t}^2$ for the case of $N_{\rm DP}^{\rm brmss}$.
The timelike form factor~$F(m_{A'}^2)$ is multiplied to take into account the mixing with $\rho$ and $\omega$ mesons.
We adopt the parametrization proposed in ref.~\cite{Faessler:2009tn}, which is based on the extended vector meson dominance approach and written as
\begin{align}
F(p_{A'}^2) = \sum_V \frac{f_V m_V^2}{m_V^2 - p_{A'}^2 - im_V\Gamma_V}~,
\label{eq:tFF}
\end{align}
where $V=\rho,\rho',\rho'',\omega,\omega',\omega''$.
The masses of $\rho$ and $\omega$ mesons are assumed to be the same for each family and set as  
$m_\rho = m_\omega = 0.77$ GeV, 
$m_\rho' = m_\omega' = 1.25$ GeV, and 
$m_\rho'' = m_\omega'' = 1.45$ GeV.
As for the decay widths, 
$\Gamma_\rho = 0.15$ GeV, $\Gamma_\omega = 0.0085$ GeV, 
$\Gamma_\rho' = \Gamma_\omega' = 0.3$ GeV, and 
$\Gamma_\rho'' = \Gamma_\omega'' = 0.5$ GeV 
are assumed.
We do not include the uncertainties of the masses and decay widths in our calculations.
With these values, it is shown in refs.~\cite{deNiverville:2016rqh,Foroughi-Abari:2021zbm} that the parametrization can fit the experimental data for 
$f_\rho = 0.616$, $f_\omega = 1.011$, 
$f_\rho' = 0.223$, $f_\omega' = -0.881$, 
$f_\rho'' = -0.339$, and $f_\omega'' = 0.369$.
It should be noted, however, that the experimental data in the timelike region are limited, in particular, near the resonance; The validity of eq.~(\ref{eq:tFF}) is still under discussion at present.

\subsection{Results}
\begin{figure}[t]
\centering
\includegraphics[width=0.48\textwidth]{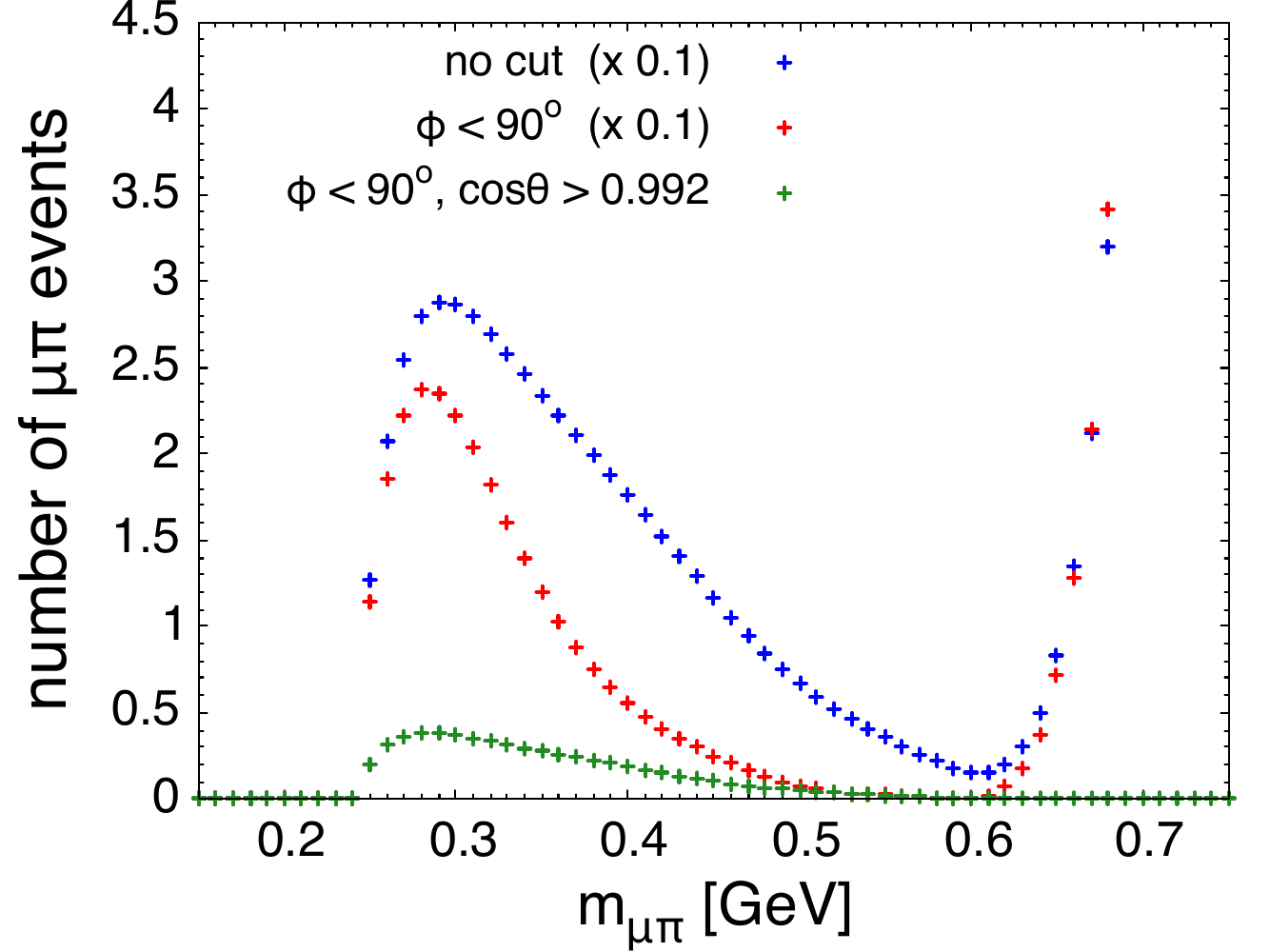}
~~~
\includegraphics[width=0.48\textwidth]{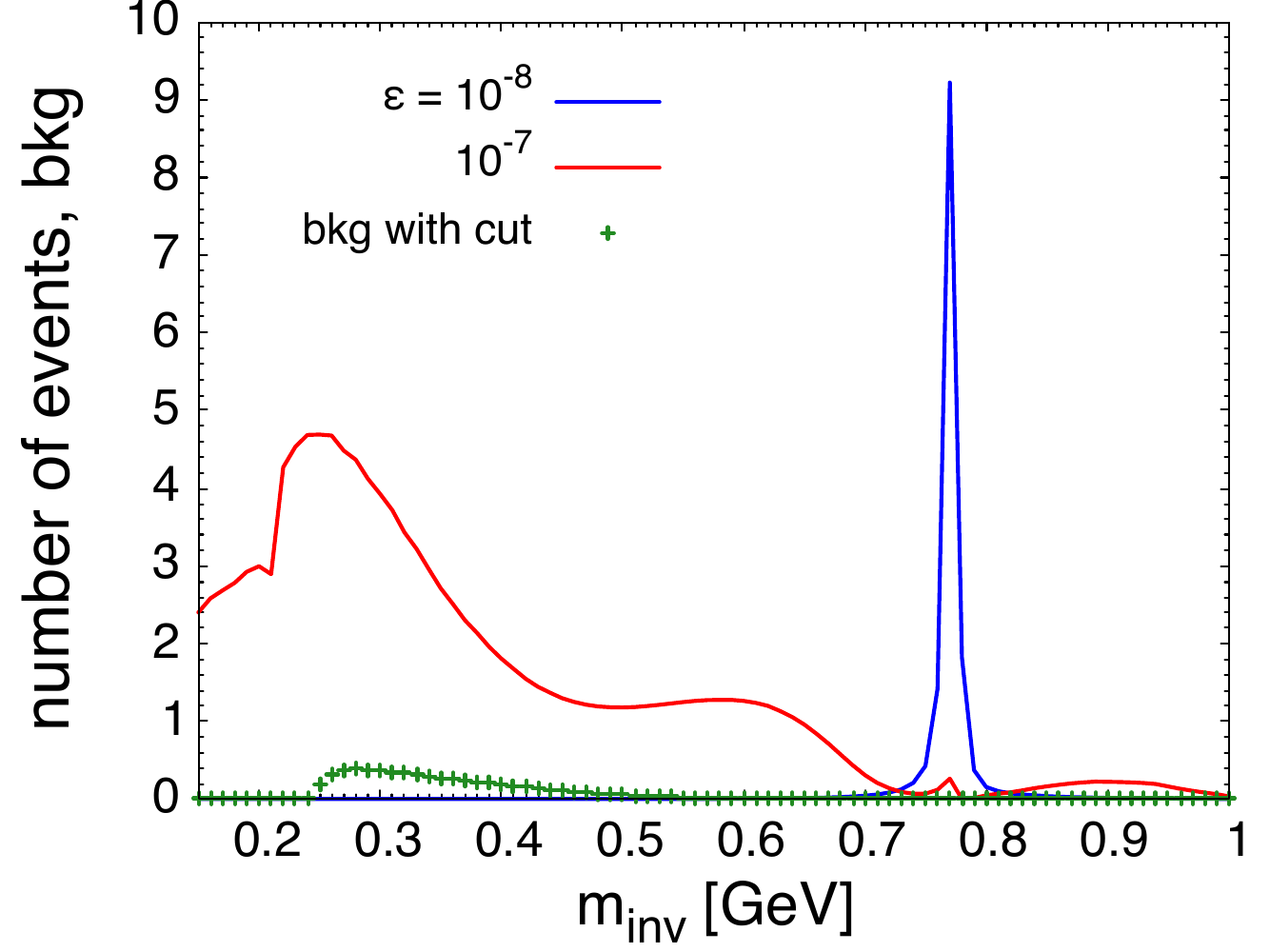}
\caption{
(left) The invariant mass distribution of $\mu^{\pm}\pi^{\mp}$ events before the event selection (blue), with the event selection of $\Delta\Phi < 90^{\circ}$ (red), and $\Delta\Phi < 90^{\circ}$ and $\cos\theta > 0.992$ (green). (right) The invariant mass distribution after the selection ($\Delta\Phi < 90^{\circ}$ and $\cos\theta > 0.992$) for the signal with $\epsilon = 10^{-8}$ (blue) and $10^{-7}$ (red) as well as background with $\mu^{\pm}\pi^{\mp}$ events (green). Both plots are scaled to $N_{\mathrm{pot}} = 3.8 \times 10^{21}$.}
\label{fig:cut}
\end{figure}
\begin{figure}[t]
\centering
~~~
\includegraphics[width=0.48\textwidth]{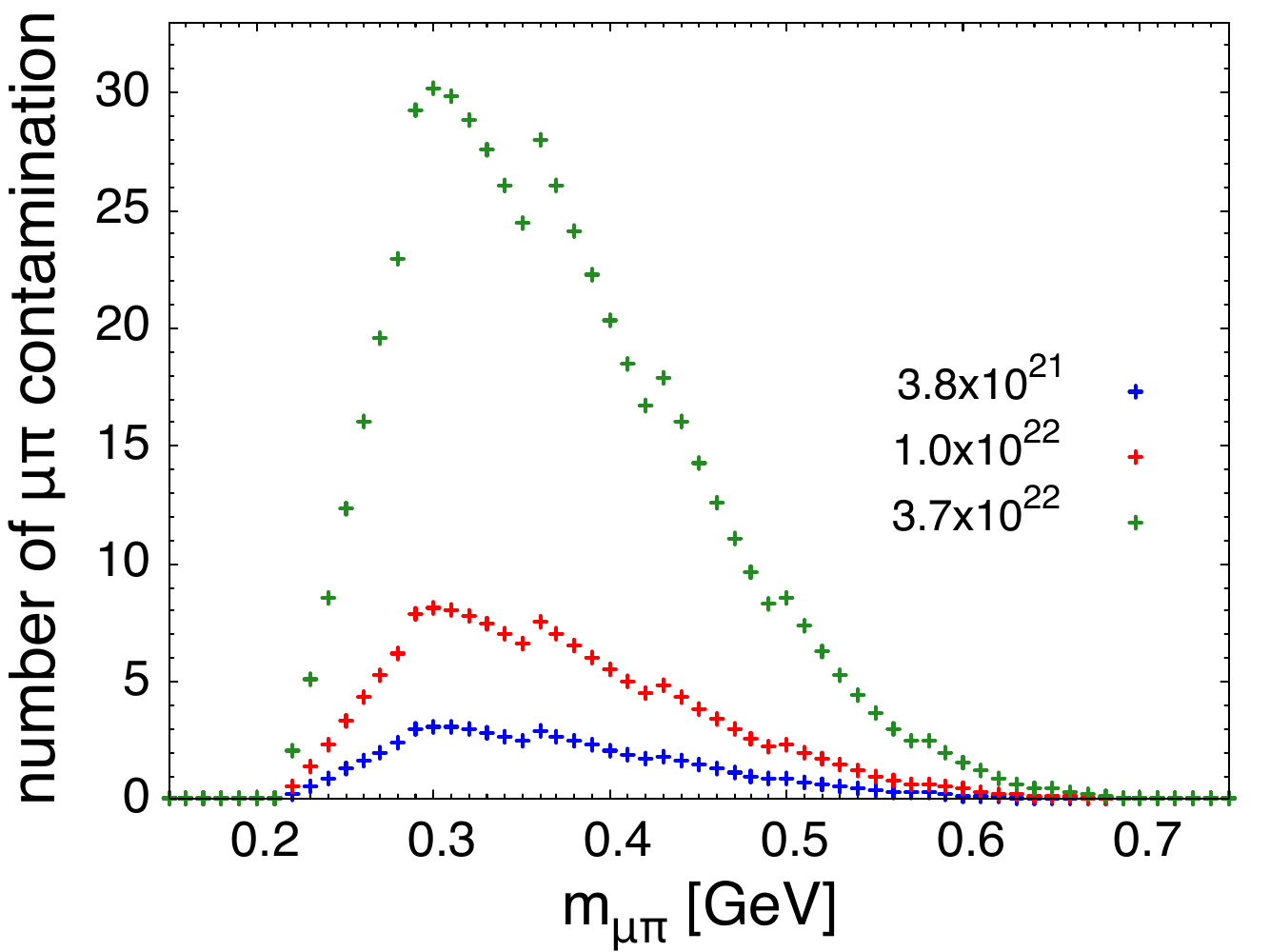}
\caption{
The number of $\mu^{\pm}\pi^{\mp}$ events that contaminate into the signal region after the mass cut with 2 $\sigma$ resolution as a function of $m_{\mu \pi}$. The blue, red and green points correspond to $N_{\mathrm{pot}} = 3.8 \times 10^{21},~1.0 \times 10^{22}$ and $3.7 \times 10^{22}$, respectively.}
\label{fig:bkg}
\end{figure}
\begin{figure}[t]
\centering
\includegraphics[width=0.7\textwidth]{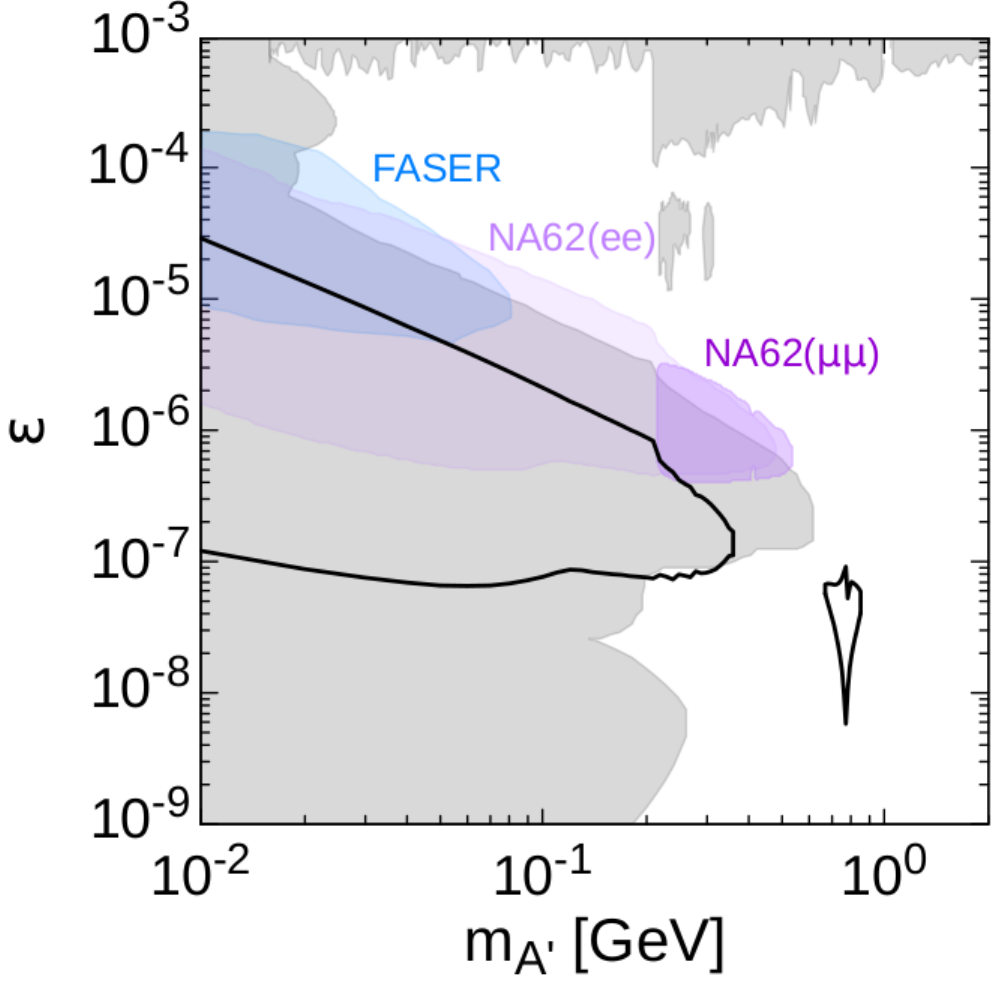}
\caption{
Exclusion regions for the dark photon model.
The regions surrounded by the black solid curves are the 95\% C.L. exclusion regions with $N_{\rm pot} = 3.8 \times 10^{21}$, that are obtained in this work.
The gray shaded areas are excluded by previous experiments and extracted from refs.~\cite{Ilten:2018crw,Kling:2021fwx}.
Also, as a reference, the recent results of 90\% C.L. exclusion regions by 
NA62~\cite{na62collaboration2023search,NA62LaThuile} and FASER~\cite{CERN-FASER-CONF-2023-001} are shown as purple and blue shaded areas, respectively.
}
\label{fig:new95CL}
\end{figure}
\begin{figure}[t]
\centering
\includegraphics[width=0.48\textwidth]{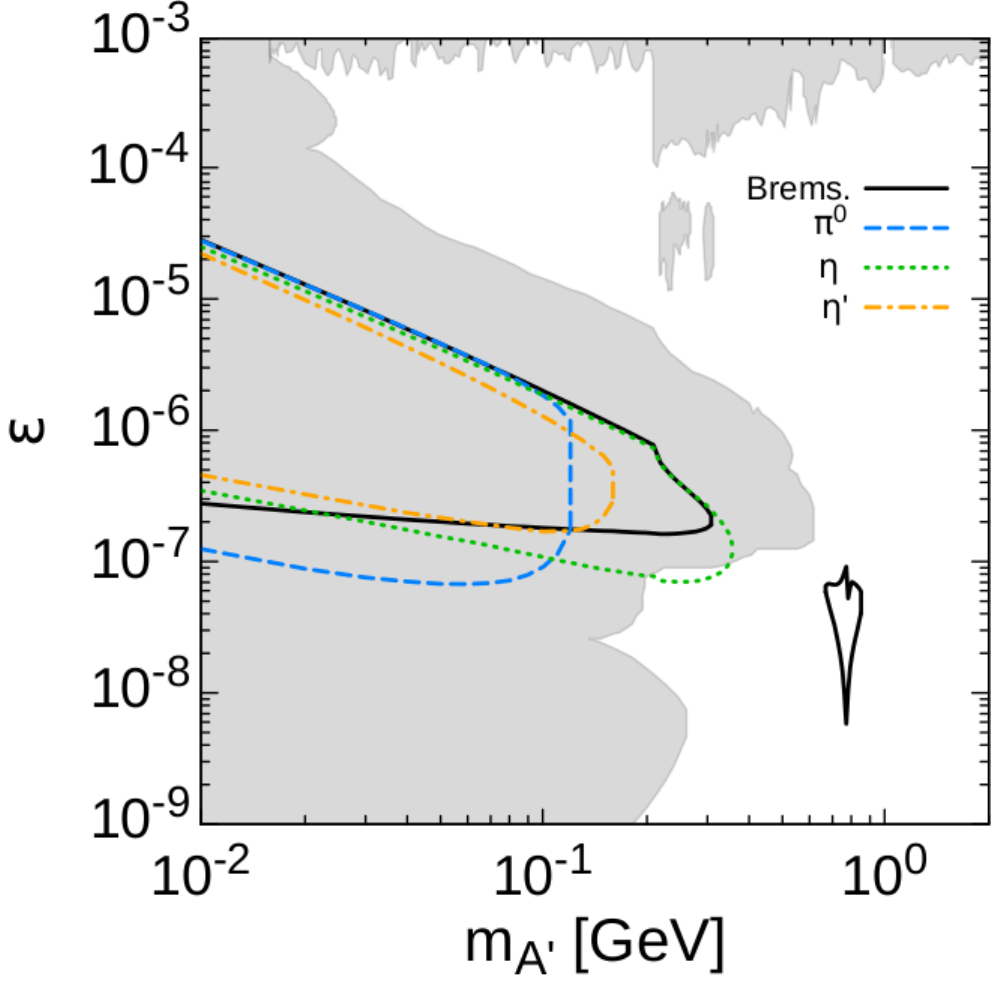}
~~~
\includegraphics[width=0.48\textwidth]{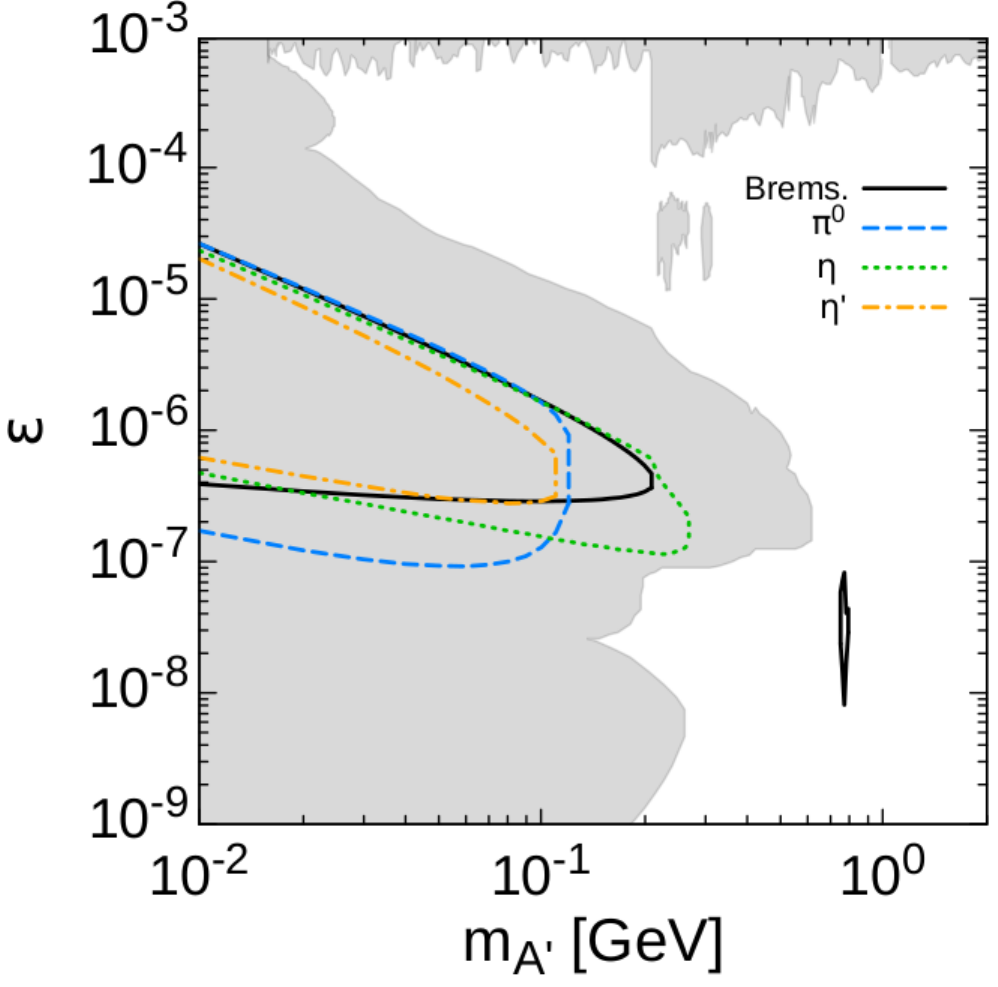}
\caption{
Left (Right): the regions predicting more than three (ten) events for each production process with $N_{\rm pot} = 3.8 \times 10^{21}$.
The black solid, blue dashed, green dotted, and orange dashed-dotted curves correspond to productions from the bremsstrahlung process, $\pi^0$ decay, $\eta$ decay, and $\eta'$ decay, respectively.
The shaded areas are the excluded regions by other experiments shown in figure~\ref{fig:new95CL}.
}
\label{fig:production}
\end{figure}
\begin{figure}[t]
\centering
\includegraphics[width=0.48\textwidth]{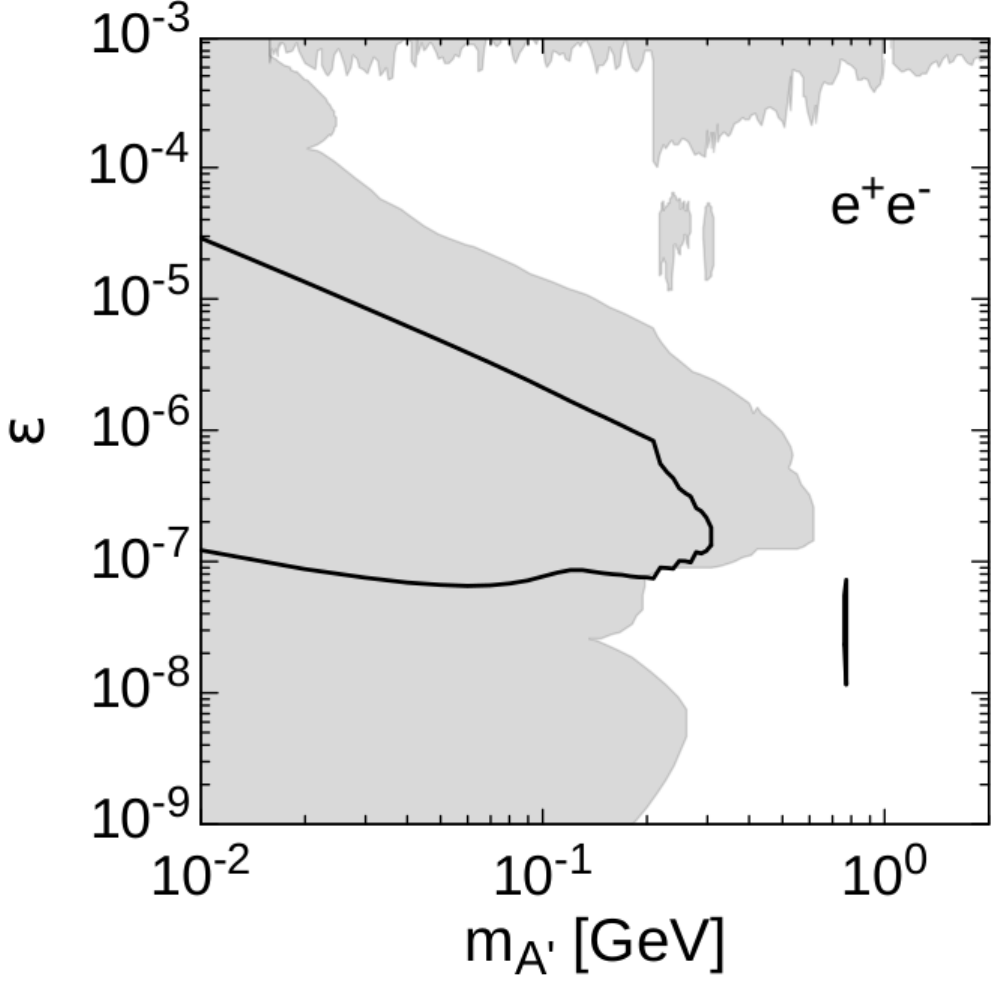}
~~~
\includegraphics[width=0.48\textwidth]{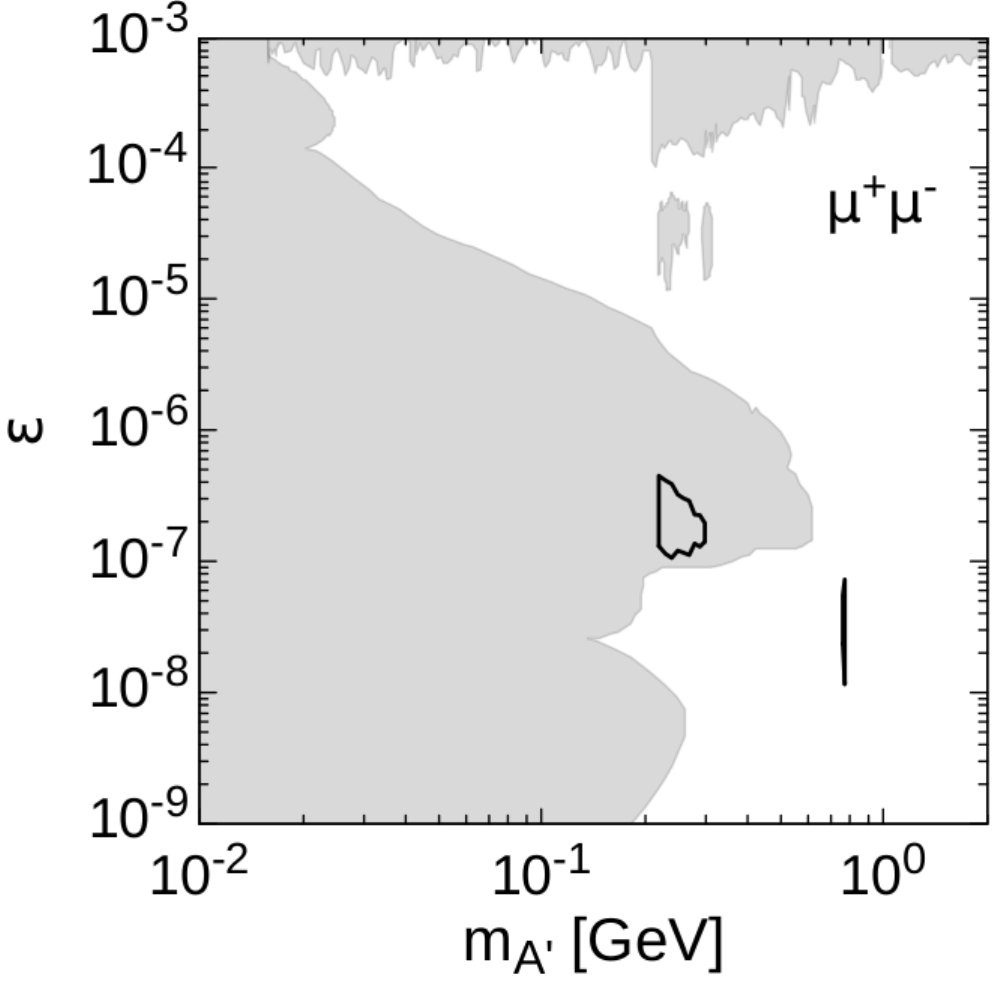}
\vspace{5mm}\\
\includegraphics[width=0.48\textwidth]{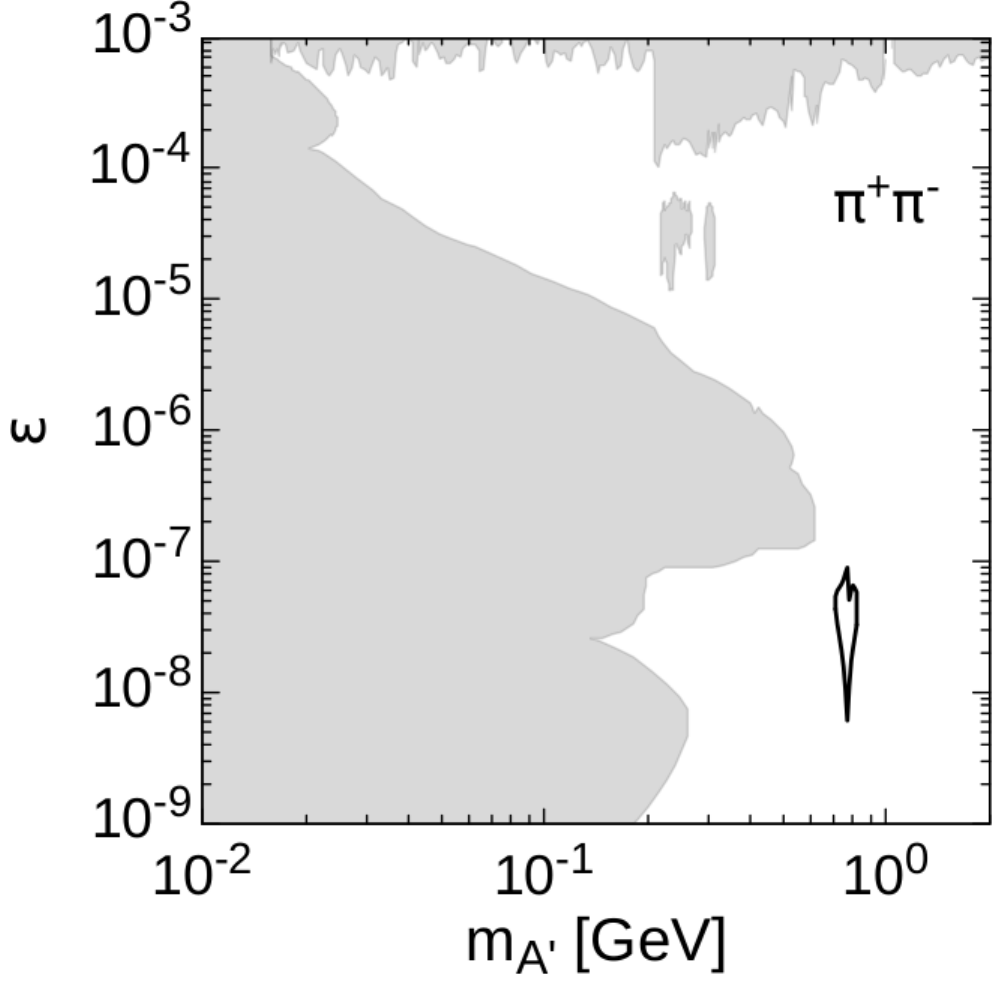}
\caption{
The 95\% C.L. exclusion regions for the $e^+ e^-$ (top left), $\mu^+ \mu^-$ (top right), and $\pi^+ \pi^-$ (bottom) final states with $N_{\rm pot} = 3.8 \times 10^{21}$.
The shaded areas are the excluded regions by other experiments shown in figure~\ref{fig:new95CL}.
}
\label{fig:final}
\end{figure}
The standard neutrino interactions with matter in the TPC become the background in this analysis. Requiring existence of two charged particles and applying the selection cuts ($\Delta\Phi < 90^{\circ}$ and $\cos\theta > 0.992$), most of the neutrino events like $\nu_{\mu} n \to \mu^{-} p$ charged current quasi elastic events can be removed. The remaining background is dominated by $\mu^{\pm}\pi^{\mp}$ final states in the charged current coherent pion production due to existence of two charged particles and difficulty to separate $\mu^{\pm}$ and $\pi^{\pm}$ with the ND280, of that we take into account in this study. The number of events with the $\mu^{\pm}\pi^{\mp}$ final states after the selection cuts which is shown in TABLE~II of ref.~\cite{T2K:2019jwa} is scaled to $N_{\rm pot}$ and the volume of TPCs which is 60\% larger in T2K-II than T2K. 
GENIE v3.0.6~\cite{Andreopoulos:2015wxa} is used to simulate the final state objects in the coherent pion production and create the invariant mass distributions with the $\mu^{\pm}\pi^{\mp}$ final states. Figure \ref{fig:cut} (left) shows the invariant mass distribution of $\mu^{\pm}\pi^{\mp}$ events before the event selection (blue), with the event selection of $\Delta\Phi < 90^{\circ}$ (red), and $\Delta\Phi < 90^{\circ}$ and $\cos\theta > 0.992$ (green) for $N_{\mathrm{pot}} = 3.8 \times 10^{21}$. It can be seen that the selection cuts efficiently reject the background. Figure \ref{fig:cut} (right) shows the invariant mass distribution after the selection ($\Delta\Phi < 90^{\circ}$ and $\cos\theta > 0.992$) for the signal with $\epsilon = 10^{-8}$ (blue) and $10^{-7}$ (red) as well as background with $\mu^{\pm}\pi^{\mp}$ events (green) for $N_{\mathrm{pot}} = 3.8 \times 10^{21}$.

As described in Sec. \ref{sec:detector}, the selection cut of $m_{\mu \pi} < 700$~MeV, that is used in ref.~\cite{T2K:2019jwa}, is not applied in this study. 
Figure \ref{fig:cut} (left) shows that the amount of the $\mu^{\pm}\pi^{\mp}$ events is not under estimated without any events at $m_{\mu \pi} > 700$\,MeV. 
Then, based on this distribution, the invariant mass cut with the 2~$\sigma$ resolution 14\% is additionally applied, considering the momentum resolution for the charged particles with ND280. 
Figure \ref{fig:bkg} shows the number of $\mu^{\pm}\pi^{\mp}$ events that contaminate the signal region after the mass cut as a function of $m_{\mu \pi}$ for $N_{\mathrm{pot}} = 3.8 \times 10^{21}$ (blue), $1.0 \times 10^{22}$ (red) and $3.7 \times 10^{22}$ (green).

We calculate 95\% C.L. exclusion regions by assuming the null observation of dark photons and rounding up the expected number of backgrounds.
The dimensions of ND280 assumed in this study are summarized in table \ref{tab:nd280} in appendix~\ref{sec:probability}.
The T2K experiment started physics data-taking in 2010 and accumulated a total of $N_{\rm pot}=3.8\times 10^{21}$ until 2021.
With this number of POT, in figure~\ref{fig:new95CL}, we derive the 95\% C.L. exclusion regions in the $m_{A'}^{} - \varepsilon$ plane.
In this figure, we also depict constraints by previous experiments as shaded areas.
As can be seen, the new exclusion regions obtained in this work place a new constraint in the parameter region around $m_{A'}^{} \simeq 0.7$~--~$0.9$\,GeV and $\varepsilon \simeq 10^{-8}$~--~$10^{-7}$.
Also, there is a marginal exclusion region around $m_{A'}^{} \simeq 0.3$~--~$0.4$\,GeV and $\varepsilon \simeq 10^{-7}$.
Compared with the past beam dump experiments, the ND280 detector is sensitive to a small kinetic mixing region. 
This is because the proton beam energy of the T2K experiment is lower and hence the produced dark photon is not boosted very much. 
As a result, the kinetic mixing must be small so that the dark photon can reach the detector. 
Although the production cross section of the dark photon is suppressed in such a situation, the high intensity of the proton beam can produce a sufficient amount of the dark photon.
In order to see the difference of the production modes, in figure~\ref{fig:production}, we show regions predicting more than three (left panel) and ten (right panel) signal events for each production process.
It is found that the new constraints are conduced from the $\eta$ meson decay and bremsstrahlung production. 
$\pi$ meson is produced more than $\eta'$ however it is too light to give a new constraint while $\eta'$ is less produced due to its mass. 
Also, in figure~\ref{fig:final}, we show the 95\% C.L. exclusion regions for each final state: $e^+ e^-$ (top left), $\mu^+ \mu^-$ (top right) and $\pi^+ \pi^-$ (bottom). 
The $\pi^+ \pi^-$ final state mainly excludes the mass region around $0.8$\,GeV. This is due to the mixing with $\rho$ and $\omega$ mesons.

\begin{figure}[t]
\centering
\includegraphics[width=0.7\textwidth]{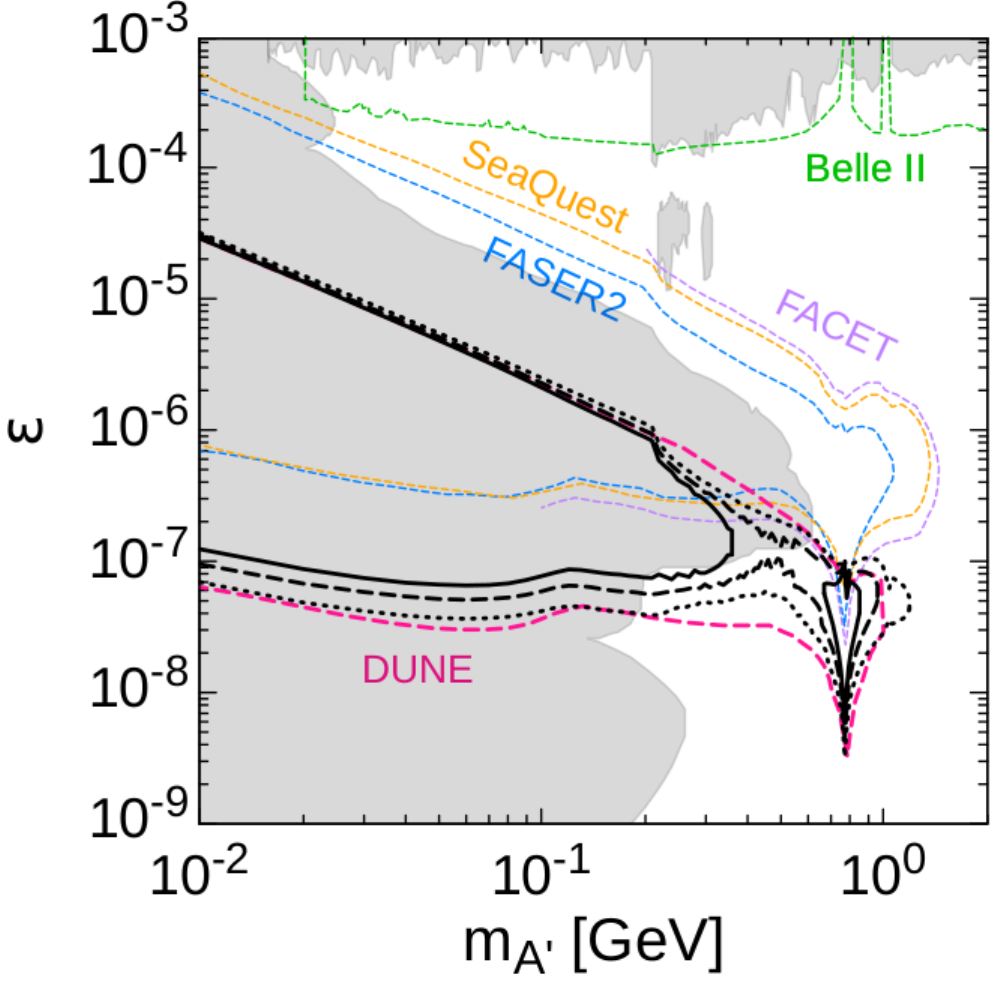}
\caption{
The excluded and expected sensitivity regions of T2K and other future experiments.
The black curves are the 95\% C.L. exclusion and expected sensitivity regions of T2K for $N_{\rm pot} = 3.8 \times 10^{21}$ (solid), $1.0 \times10^{22}$ (dashed) and $3.7 \times 10^{22}$ (dotted). 
The colored dashed curves represent the expected sensitivity reaches of Belle II (green) \cite{Belle-II:2018jsg}, DUNE (red) \cite{Berryman:2019dme}, FACET (purple) \cite{Cerci:2021nlb}, FASER2 (blue) \cite{FASER:2018eoc} and SeaQuest (orange) \cite{Berlin:2018pwi}.
The shaded areas are the excluded regions by other experiments shown in figure~\ref{fig:new95CL}.
}
\label{fig:future95CL}
\end{figure}
The ND280 detector is upgraded in 2023 as described in section \ref{sec:detector} and will accumulate $N_{\rm pot} = 1.0\times 10^{22}$ data by the end of 2027.
Furthermore, after 2027, the experiment with Hyper-Kamiokande (HK) will start, and $N_{\rm pot} = 3.7 \times 10^{22}$ is expected in ten years.
In figure~\ref{fig:future95CL}, we show the expected sensitivity regions for these future projects and compare them with the projections of other future experiments.
As can be seen from the figure, the sensitivity regions of T2K are very similar to that of the DUNE experiment \cite{Berryman:2019dme}, and they are sensitive for smaller $\varepsilon$ compared with those of collider and beam dump experiments.
It is interesting that the long-baseline neutrino oscillation experiments complement collider and beam dump experiments, and open up a new avenue in the exclusion plot of the dark photon model.

\begin{figure}[t]
\centering
\includegraphics[width=0.7\textwidth]{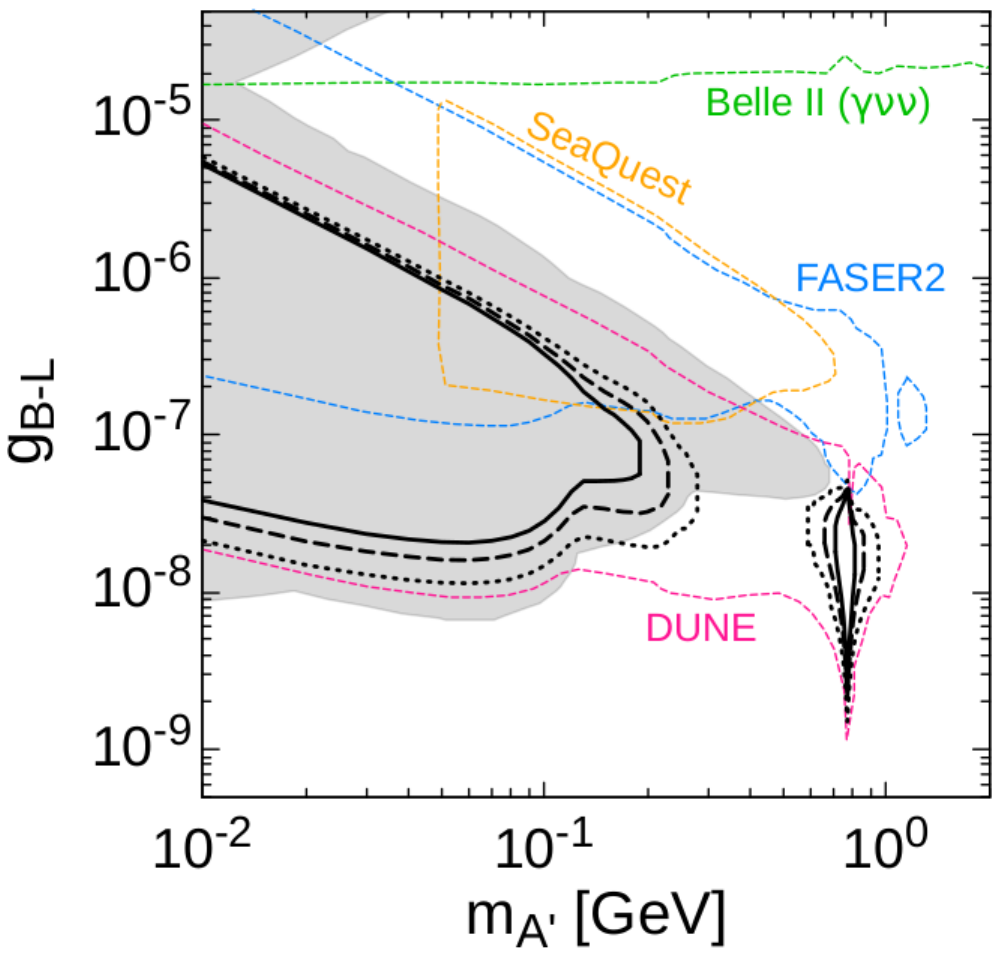}
\caption{
The 95\% C.L. exclusion region and expected sensitivity regions of T2K for the U(1)$_{B-L}$ model.
The black solid, dashed and dotted curves correspond to $N_{\rm pot} = 3.8 \times 10^{21}$, $1.0 \times10^{22}$, and $3.7 \times 10^{22}$, respectively.
The gray shaded areas are excluded by previous experiments and extracted from refs.~\cite{Ilten:2018crw,Kling:2021fwx}.
The colored dashed curves represent the expected sensitivity reaches of other experiments: Belle II (green) and SeaQuest (orange) are taken from ref. \cite{Bauer:2018onh}, DUNE (red) from ref. \cite{Dev:2021qjj}, and FASER2 (blue) from ref. \cite{Feng:2022inv}.
}
\label{fig:B-L}
\end{figure}
Before closing this study, we demonstrate the applicability of our calculation to other light gauge bosons.
As an example, we here consider a U(1)$_{B-L}$ gauge boson having the following Lagrangian:
\begin{align}
{\cal L}_{\rm B-L}^{} = 
- g_{B-L}^{} \sum_{f} Q_f~ \bar{f}\gamma^\mu f~ A_{\mu} '
+ \frac{1}{2}m_{A'}^{2} A_\mu' A'^{\mu} ~,
\label{eq:B-L}
\end{align}
where $g_{B-L}^{}$ stands for the new gauge coupling constant. 
The sum runs over all the SM fermions with the U(1)$_{B-L}$ charge $Q_{f}=1/3$ for all the quarks, while $Q_{f}=-1$ for leptons.
In accordance with refs. \cite{Tulin:2014tya,deNiverville:2016rqh,Ilten:2018crw}, we multiply $g_{B-L}^2/(\varepsilon e)^2$ and $g_{B-L}^2/(2\varepsilon e)^2$ by ${\rm BR}(\pi^0 \rightarrow A'\gamma)$ and ${\rm BR}(\eta \rightarrow A'\gamma)$ in eq. (\ref{eq:BRprod}), respectively
\footnote{We take a limit of $m_{A'} \ll m_{\rho,\omega,\phi}$ and assume the Breit-Wigner functions are equal to unity.}, 
while we ignore production from $\eta'$ mesons since it is suppressed in comparison with $\pi^0$ and $\eta$ for the case of U(1)$_{B-L}$.
Similarly, we multiply $g_{B-L}^2/(\varepsilon e)^2$ by both the partial decay width in eq. (\ref{eq:Gll}) and the number of events from the bremsstrahlung production in eq. (\ref{eq:Nbrmss}).
Here, it should be stressed that U(1)$_{B-L}$ gauge bosons can decay into neutrinos; We assume three generations of massless Majorana neutrinos~\footnote{
For Dirac neutrinos, the partial decay width is multiplied by 2.}
with $\Gamma_{\nu\nu} = 3 \times g_{B-L}^2 m_{A'}/(24\pi)$.
Note also that U(1)$_{B-L}$ gauge bosons do not mix with $\rho$ mesons because of the universal charge assignment for quarks.
As a result, the decay branching ratio of $A' \rightarrow \pi^+\pi^-$ is suppressed to be negligible level.
Thus, we ignore $A' \rightarrow \pi^+\pi^-$ and, also, turn off the resonant mixing with $\rho,\rho',\rho''$ mesons in the timelike form factor included in eq. (\ref{eq:Nbrmss}).
By incorporating only $e^+ e^-$ and $\mu^+ \mu^-$ final states, in figure \ref{fig:B-L}, we derive the 95\% C.L. exclusion region and expected sensitivity regions of T2K for U(1)$_{B-L}$.
In comparison with the case of dark photons, sensitivity regions disappear around 0.2 GeV - 0.6 GeV because of the lack of the resonant mixing with $\rho$ mesons.

\section{Summary}
\label{sec:summary}
We have studied the constraints on the dark photon and U(1)$_{B-L}$ gauge boson from T2K off-axis ND280 detector. 
We have simulated the dark photon production from light meson decays and proton bremsstrahlung assuming $3.8 \times 10^{21}$ POT, corresponding to ten-years operation of T2K. 
It is found that the unexplored parameter region is excluded due to no observation of two electron, muon and pion tracks as shown in figures \ref{fig:new95CL} and \ref{fig:B-L}, for dark photon and U(1)$_{B-L}$ gauge boson, respectively. 
We also analyzed the expected sensitivity for the future upgrade of the ND280 detector and proton beam line. 
The results are shown in figures \ref{fig:future95CL} and \ref{fig:B-L}. 
One can see that the upgraded T2K experiment will complement the searches by other future experiments such as FASER2, SeaQuest, and FACET.

The analyses given in this paper can be applied to other new particles beyond the SM, for instance, axion-like particles and light scalar bosons. 
These particles also can be produced from the meson decays and proton bremsstrahlung.
When one considers a light scalar boson as the origin of the dark photon mass, the decay of such a scalar boson will give significant contributions to the production of the light gauge boson (see, for instance, ref.~\cite{Araki:2020wkq}). 
We leave these topics for our future works.

\begin{acknowledgements}
The authors thank Ken Sakashita for the discussion on the T2K experiment and the ND280 detector as well as Tomoko Ariga and Tomohiro Inada for technical advice to evaluate backgrounds from neutrino interactions. 
This work was supported by JSPS KAKENHI Grant Numbers JP21K20365 [KA], JP23K13097 [KA], JP23H01189 [HO, TA, TS, YT], JP20K04004 [YT], JP21H00082 [YT], JP20H01919 [HO, YT], JP18H01210 [TS], JP18K03651 [TS], and MEXT KAKENHI Grant Numbers JP18H05543 [TS], JP22K03622 [TS].
\end{acknowledgements}

\newpage

\appendix
\renewcommand{\thesubsection}{\Alph{section}.\arabic{subsection}}
\section{Setup of detector and detection probability}
\label{sec:probability}
\begin{table}[t]
\begin{tabular}{|c|c|c|c|c|c|} \hline
 ~~~width (m)~~~ & ~~~height (m)~~~ & ~~~depth (m)~~~ & ~~~$L_{\rm min}$ (m)~~~ & ~~~$L_{\rm max}$ (m)~~~ & ~~~$\theta_{\rm ND}$ (deg)~~~ \\ \hline \hline 
2.4 & 2.4 & 5.8 & 280.1 & 285.9 & 2.0 \\ \hline
\end{tabular}
\caption{
The dimensions and location of the ND280 detector.
}
\label{tab:nd280}
\end{table}

\begin{figure}[h]
\centering
\includegraphics[width=0.6\textwidth]{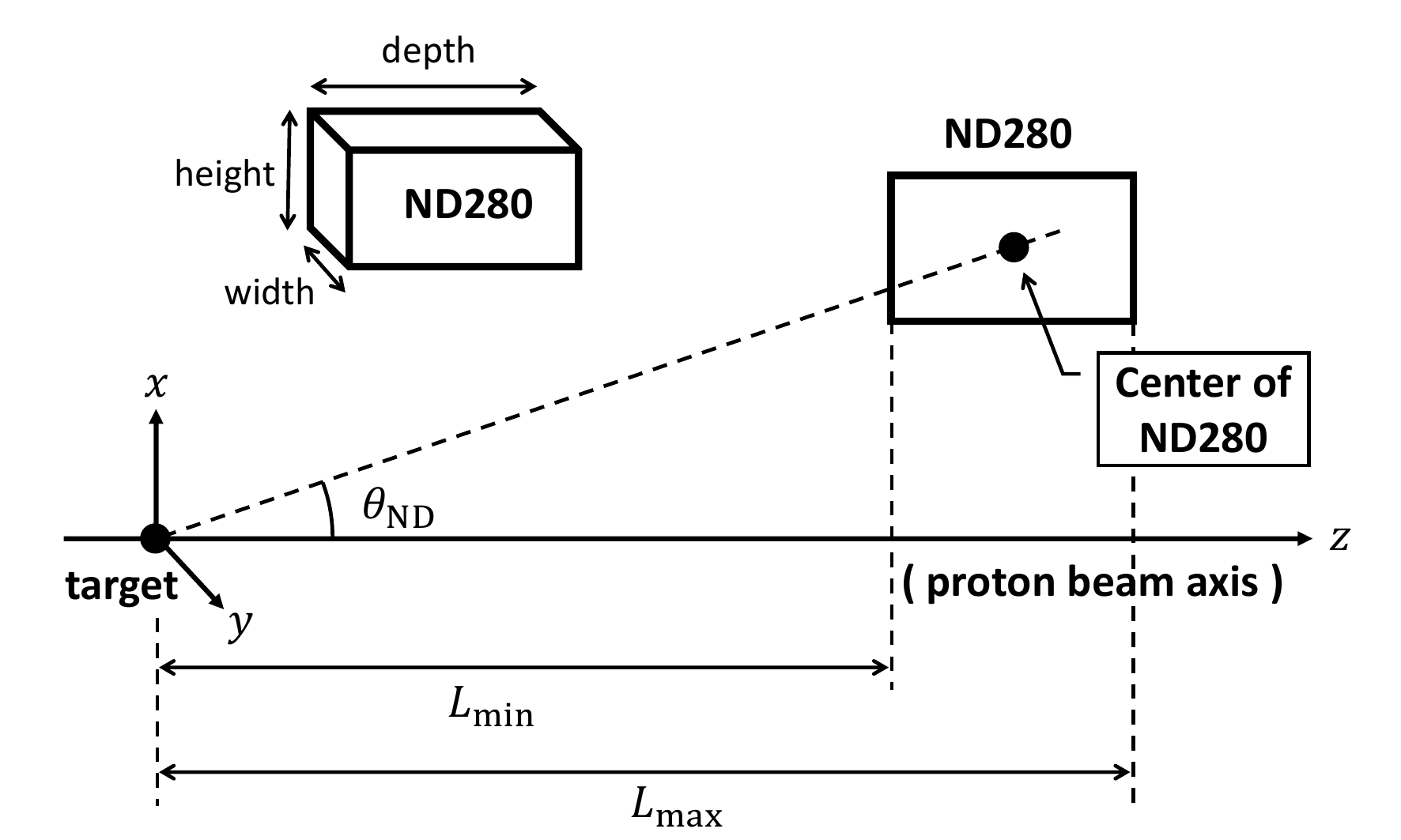}
\caption{
A schematic view of the location of ND280.
}
\label{fig:nd280dim}
\end{figure}
\begin{figure}[h]
\centering
\includegraphics[width=0.8\textwidth]{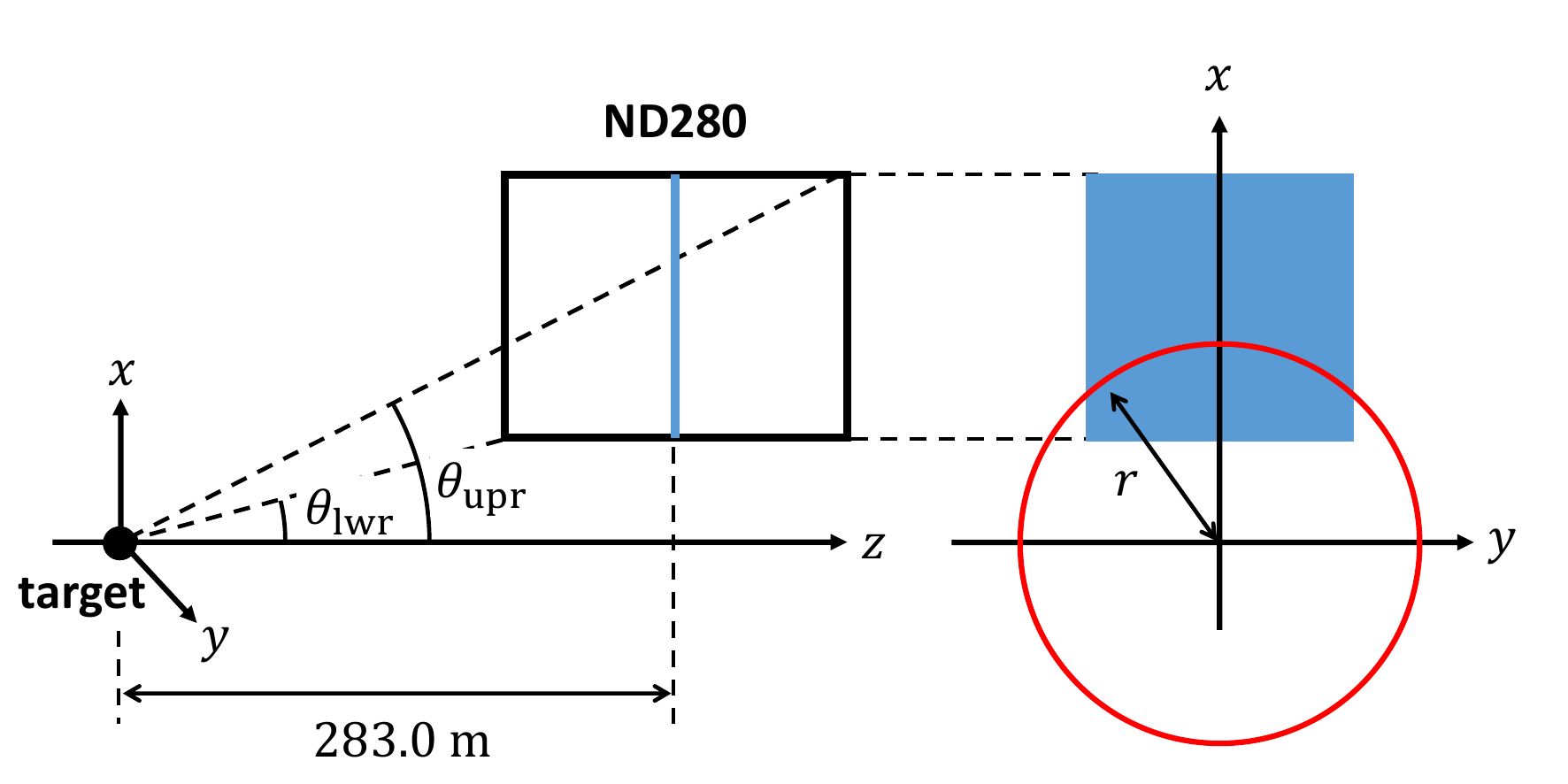}
\caption{
The angular acceptance of ND280.
}
\label{fig:nd280acp}
\end{figure}

In our numerical calculations, we assume that the ND280 detector is a rectangular with dimensions of $2.4$ m $\times$ $2.4$ m $\times$ $5.8$\,m. 
We also assume that the target is point like, the detector is located $280.1$\,m away from the target, positioned in the direction of $2^\circ$\footnote{This angle at the target corresponds to the direction of $2.5^\circ$ at the mean decay point of charged pions.} from the beam axis, and installed parallel to the beam axis.
The dimensions and location of ND280 are summarized in table~\ref{tab:nd280} and figure~\ref{fig:nd280dim}.

We formulate a probability that a dark photon decays inside the detector volume as
\begin{align}
{\cal P}^{\rm det}(|{\bm p}_{A'}^{}|, \theta_{A'})
=\frac{1}{d_z}
\int_{L_{\rm min}}^{L_{\rm max}} \dd z~ e^{-\frac{z}{d_z}}~
P(\theta_{A'}, z) ~\times~ 
\Theta(\theta_{\rm upr}-\theta_{A'}) \times
\Theta(\theta_{A'}-\theta_{\rm lwr})~,
\label{eq:Prob}
\end{align}
where $z$ is a $z$-coordinate at which the dark photon decays, and the momentum of the dark photon and its angle with respect to the beam axis are denoted as ${\bm p}_{A'}^{}$ and $\theta_{A'}$, respectively.
The decay length of a dark photon is defined as 
\begin{align}
d_z = \frac{c\hbar}{\Gamma_{\rm total}} \beta\gamma \cos\theta_{A'}~,
\end{align}
where $c$ is the speed of light, $\hbar$ is the reduced Planck constant, $\beta\gamma = |{\bm p}_{A'}^{}|/m_{A'}^{}$ is the Lorentz factor, and $\Gamma_{\rm total}$ is the total decay width of dark photons.

The Heaviside step functions, $\Theta$, in eq.~\eqref{eq:Prob} restrict the range of $\theta_{A'}$.
The upper and lower limit, $\theta_{\rm upr}$ and $\theta_{\rm lwr}$, respectively, are set as depicted in the left panel of figure~\ref{fig:nd280acp} and are given by
\begin{align}
\theta_{\rm lwr} = \arctan{\left( \frac{h-2.4\,{\rm m}/2}{280.1\,{\rm m}} \right)},~~~~
\theta_{\rm upr} = \arctan{\left( \frac{h+2.4\,{\rm m}/2}{285.9\,{\rm m}} \right)}~,
\label{eq:th_limit}
\end{align}
where $h=(280.1+5.8/2)\,{\rm m} \times \tan 2^\circ$.

The function $P(\theta_{A'}, z)$ restricts the angle between the $x$ axis and the transverse component of a dark photon momentum; we estimate it by calculating how long does the circumference drawn by dark photons having $\theta_{A'}$ and decaying at $z$ overlap the detector volume, see the right panel of figure~\ref{fig:nd280acp}.
Depending on $r=z\tan \theta_{A'}$, the form of $P(\theta_{A'}, z)$ is divided into the following three cases: 
\begin{itemize}
\item[(i)] $h-\frac{2.4\,{\rm m}}{2} < r < \sqrt{ \left(h-\frac{2.4\,{\rm m}}{2}\right)^2 + \left(\frac{2.4\,{\rm m}}{2}\right)^2 }$
\begin{align}
P(\theta_{A'}, z) = \frac{1}{\pi} \arccos\left( \frac{h-2.4\,{\rm m}/2}{r} \right)~,
\end{align}

\item[(ii)] $\sqrt{ \left(h-\frac{2.4\,{\rm m}}{2}\right)^2 + \left(\frac{2.4\,{\rm m}}{2}\right)^2 } < r < h+\frac{2.4\,{\rm m}}{2}$
\begin{align}
P(\theta_{A'}, z) = \frac{1}{\pi} \arcsin\left( \frac{2.4\,{\rm m}}{2r} \right)~,
\label{eq:Pzt2}
\end{align}

\item[(iii)] $h+\frac{2.4\,{\rm m}}{2} < r < \sqrt{ \left(h+\frac{2.4\,{\rm m}}{2}\right)^2 + \left(\frac{2.4\,{\rm m}}{2}\right)^2 }$
\begin{align}
P(\theta_{A'}, z) = \frac{1}{\pi}
\left\{ 
 \arcsin\left( \frac{2.4\,{\rm m}}{2r} \right)
-\arccos\left( \frac{h+2.4\,{\rm m}/2}{r} \right) 
\right\}~.
\end{align}
\end{itemize}
Nevertheless, in our numerical calculations, we drop the $z$ dependence in $P(\theta_{A'}, z)$ by fixing $r$ at the center of the detector, i.e., $r=283\,{\rm m} \times \tan \theta_{A'}$. This can be validated by the fact that the detector depth is much shorter than the distance between the target and detector. 
In this case, the $z$ integral in eq.~\eqref{eq:Prob} can be analytically done, only case (ii) applies, and the probability is simplifyed to be
\begin{align}
& {\cal P}^{\rm det}(|{\bm p}_{A'}|, \theta_{A'})
=
\left( e^{-L_{\rm min}/d_z} - e^{-L_{\rm max}/d_z} \right)
\times \frac{1}{\pi} \arcsin\left( \frac{1.2\,{\rm m}}{283\,{\rm m} \times \tan\theta_{A'}} \right)
\nonumber\\
& \hspace{3cm}
\times
\Theta(\theta_{\rm upr}-\theta_{A'}) \times
\Theta(\theta_{A'}-\theta_{\rm lwr})~,
\label{eq:Paprx}
\end{align}
where $\theta_{\rm upr}$ and $\theta_{\rm lwr}$ are given in eq.~\eqref{eq:th_limit}.

\bibliographystyle{utphys28mod}
{\small 
\bibliography{ref}
}

\end{document}